\documentclass{article}    
\usepackage{epsfig}      
\usepackage{url}
\newcommand{\rsub}[1]{\mbox{\tiny #1}}
\newcommand{\e}[1]{\begin{equation}\label{#1}}
\newcommand{\ee}{\end{equation}}

\title{An ionically based mapping model with memory \\
for cardiac restitution  }

\author{
David G. Schaeffer$^{1,4}$, John W. Cain$^{1}$, Daniel J. Gauthier$^{2,3,4}$, 
Soma S. Kalb$^{3}$, \\ 
 \\
Wanda Krassowska$^{3,4}$, Robert A. Oliver$^{3}$, and 
Elena G. Tolkacheva$^{2}$ \\
 \\
\small Departments of $^{1}$Mathematics, $^{2}$Physics, and 
$^{3}$Biomedical Engineering\\ 
\small and $^{4}$Center for Nonlinear and Complex Systems,\\
\small Duke University, Durham, North Carolina 27708\bigskip , USA}

\begin{document}
\maketitle

\bigskip

{\small
\begin{center}
{\sc Abstract}
\end{center}
Many features of the sequence of action potentials produced by repeated 
stimulation of a cardiac patch can be modeled by a 1D mapping, but not the 
full behavior observed in the restitution portrait: in particular, not 
(i)~distinct slopes for dynamic and S1-S2 restitution (rate dependence) and 
not (ii)~long transients in the approach to steady state (accomodation).
To address these shortcomings, \emph{ad hoc} 2D mappings, where the second 
variable is a ``memory'' variable, have been proposed; it seems that these 
models exhibit some, but not all, of the relevant behavior.
In this paper we introduce a new 2D mapping and determine a set of parameters 
for it that gives a rather accurate description of the full restitution 
portrait found for one animal.
The fitting procedure is straightforward and can easily be applied to obtain 
a mathematical model for data from other experiments, including experiments 
on different species.  Uniqueness of the parameter choice is also discussed.

The changes in the mapping, compared to previous models, result from 
requiring that the mapping can be derived as an asymptotic limit of a simple 
ionic model.
Among other benefits, one can interpret the parameters in the mapping in 
terms of the ionic model.
The ionic model is an extension of a two-current model that adds a third 
dependent variable, a generalized concentration.
The simplicity of the ionic model and the physiological basis for the 
mapping contribute to the usefulness of these ideas for describing 
restitution data in a variety of contexts.
}

\bigskip

\section{Introduction}

Sudden cardiac death kills half of all people who die of heart disease, the 
Number One cause of death in the United States, accounting for more than 
400,000 deaths each year \cite{web}.
As part of a larger effort to understand and prevent arrythmias that lead to 
death, much work has been devoted to studying the resitution properties of 
cardiac tissue under repeated stimulation, especially rapid stimulation 
\cite{hall-BG, banville-G, fox-GB, cherry-F}.
The \emph{restitution portrait}, as is obtained in the recently proposed 
perturbed downsweep protocol \cite{kalb-DTIKG}, combines data about both 
dynamic restitution and S1-S2 restitution, as well as information about the 
transients in the approach to steady state.
(These terms are defined 
in the Appendix.)
Because the restitution portrait includes such a wealth of information, 
experimental data from this protocol provide a demanding test of mathematical 
models for restitution.
Specifically, restitution portraits obtained from small pieces of frog 
myocardium \cite{kalb-DTIKG} show two noteworthy behaviors:
\begin{itemize}
\item[1.] Under rapid pacing but still exhibiting a 1:1 response, the 
dynamic restitution curve is typically quite steep, having slope 3 or even 
greater, while the S1-S2 restitution curve is rather shallow, having a slope 
on the order of 0.5.
\item[2.] The approach to steady state is very slow, with a time constant on 
the order of 20--30 sec.  The approach, which is monotonic, occurs along a 
smooth curve that is distinct from either restitution curve.  This behavior 
is observed over the entire range of pacing intervals leading to a 1:1 
response.
\end{itemize}
These behaviors illustrate phenomena in cardiac electrophysiology variously 
known as \emph{rate dependence}, \emph{memory}, and \emph{accomodation}.

Consider a typical experiment of producing a sequence of action potentials 
by the application of periodic stimuli to a small patch of cardiac 
tissue\footnote{Of course the heart is a spatially extended structure.  In 
performing experiments with a small piece of tissue, one hopes that 
propagation effects can be eliminated, in order to isolate a simpler 
problem.  In any event, in this paper we consider models without propagation.}.
Let $B$ (mnemonic for basic cycle length) be the period of the stimulus 
application, and let $A_n$ be the duration of the $n^{th}$ action potential.
Nolasco and Dahlen \cite{nolasco-D} proposed to describe the sequence $A_n$ 
approximately by a recursion relation
\e{1d-recursion}
	A_{n+1} = G( B - A_n )
\ee
for an appropriate function $G$.
Of course $B - A_n $ is the diastolic interval $D_n$ following the $n^{th}$ 
action potential.

Although the introduction of 1D mapping models was a decisive advance in our 
understanding of restitution, such models cannot exhibit Behaviors 1 or 2 
above; indeed, for a mapping of the form (\ref{1d-recursion}), dynamic 
restitution, S1-S2 restitution, and the approach to steady state all fall 
on the \emph{same} curve.
In an effort to model these behaviors, various authors 
\cite{chialvo-MJ, fox-BG} have introduced \emph{ad hoc} 2D mapping models, 
one variable being $A_n$ and the other a memory variable.
In unpublished analysis we have shown that parameters may be chosen in these 
models to obtain some of the behavior in \cite{kalb-DTIKG}, but we believe 
that no single set of parameter values can produce the complete range of
behavior.

In this paper we introduce a 2D map that 
\begin{itemize}
\item[(i)] fits all the dynamic restitution data, all the S1-S2 restitution 
data, and most of the transient data from the restitution portrait 
\cite{kalb-DTIKG} that involves a 1:1 response (in particular, exhibiting 
both Behavior~1 and Behavior~2)
\end{itemize}
and moreover
\begin{itemize}
\item[(ii)] can be derived with asymptotic analysis from a simple ionic model.
\end{itemize}
Regarding (i), determining the best parameter fit is a straightforward 
process, and it may be applied to model other restitution experiments with 
different animals.
Regarding (ii), in this ionic model, memory is associated with the 
accumulation over time of ions in the cell.
The ionic model suggests some subtle changes in the mapping compared to 
previous models of this type \cite{chialvo-MJ, fox-BG}, and these changes 
led to a good fit with experiment.

\section{The mapping model and its ionic basis}

Like \cite{chialvo-MJ, fox-BG}, our mapping contains two variables---one is 
$A_n$ as above; the other, which will be denoted by $c_n$, specifies the 
ionic concentration in the underlying ODE at the start of the $n^{th}$ 
action potential.
We suppose that, under periodic stimulation (say with period $B$), these 
two variables evolve according to the iteration
\e{A}
	A_{n+1} = G(D_n) + \Phi(c_{n+1})
\ee
\e{c}
	c_{n+1} = (c_n + \epsilon) e^{-B/\tau_{\rsub{pump}}}
\ee
where
\e{G}
	G(D) = A_{\rsub{max}} 
	+ \tau_{\rsub{fclose}} \ln \left\{ 1- \alpha e^{-D/\tau_{\rsub{open}}} \right\} 
\ee
and
\e{Phi}
	\Phi(c) = \tau_{\rsub{sclose}} \ln \left\{ \frac{e^{-c} + \beta}{1+\beta} \right\} .
\ee
In (\ref{A}--\ref{Phi}), 
$A_{\rsub{max}}, \tau_{\rsub{fclose}}, \tau_{\rsub{sclose}}, \tau_{\rsub{open}}$, and $\tau_{\rsub{pump}}$ are parameters with the dimensions of time, and 
$\epsilon, \alpha$, and $\beta$ are dimensionless parameters.
As we discuss below, this mapping was derived as an asymptotic limit of 
an ionic model, and the constants in (\ref{G}, \ref{Phi}) refer to 
constants in the ionic model.
Thus, by relating the mapping to the ionic model, we will be able to 
understand the significance of these constants, as well as explain the 
mnemonics of their subscripts.
This physiological connection enhances the utility of the mapping in modeling 
various experiments.

The mapping (\ref{A}, \ref{c}) must be supplemented with the restriction 
that no action potential is produced unless the diastolic interval (DI) is 
long enough.
Physically, if the DI is too short, the system will skip a beat and may 
jump to a 2:1 response.
Mathematically, if the DI is less than $\tau_{\rsub{open}} \ln \alpha$, 
then (\ref{G}) is not even defined.
The precise lower bound for the DI, which depends on the concentration $c$, 
cannot be derived without analysis of the ionic model.
These somewhat technical issues will be addressed in a future publication.

The ionic model underlying (\ref{A}, \ref{c}) is a modest extension of the 
two-current model \cite{karma, mitchell-S} in that: (a)~the equations for 
the two variables in the two-current model, the voltage $v$ and the gate $h$, 
are slightly more complicated and (b)~a third variable, the 
concentration $c$, is added to the system.  Specifically:

(i) In the modified equation for $v$,
\e{dvdt-new}
	\frac{dv}{dt}= - J_{\rsub{in}}(v,h,c) - J_{\rsub{out}}(v) ,
\ee
the outward current is still given by $J_{\rsub{out}}(v) = v /  \tau_{\rsub{out}}$, but the inward current is now the sum of  concentration-independent and  concentration-dependent parts
\e{J-new}
	J_{\rsub{in}}(v,h,c)= - \frac{h}{\tau_{\rsub{in}}}
	\left\{ \phi_{\rsub{ci}}(v) + e^{-c} \phi_{\rsub{cd}}(v) \right\}.
\ee
The response of the model to stimulation is not very sensitive to the exact form of the functions
$\phi_{\rsub{ci}}(v)$ and $\phi_{\rsub{cd}}(v)$, and we do not specify these functions here.

(ii) In the modified equation for $h$,
\e{dhdt-new}
	\frac{dh}{dt}=\left\{ \begin{array}{lll}
	(1-h)/ \tau_{\rsub{open}} & \mbox{if $v<v_{\rsub{crit}}$} \\
         & \\
	-h/ T_{\rsub{close}}(v) & \mbox{if $v>v_{\rsub{crit}}$} ,
\end{array} 
\right.
\ee
the time constant for the closing of the gate depends on the voltage $v$.
The precise form of $T_{\rsub{close}}(v)$ is less important than the fact that two different time-scale parameters derive from the closing of the gate, say 

\indent $\tau_{\rsub{fclose}}$:  \quad  fast closing for $v$ near the top of its range, and \\
\indent $\tau_{\rsub{sclose}}$:  \quad  slower closing for smaller $v$ 
	(but still greater than $v_{\rsub{crit}}$).

\medskip						

(iii) The equation\footnote{Equation (\ref{dcdt}) appears to be nonautonomous.  There is an autonomous form of the equation in which $I(t)$ is expressed in terms of the voltage $v(t)$.  However, $I(t)$ is a function of $dv/dt$ rather than $v$ itself, and we prefer the notation in (\ref{dcdt}).}
for $c$ may be written
\e{dcdt}
	\frac{dc}{dt} = I(t) - c / \tau_{\rsub{pump}}.
\ee
Thus, the concentration is determined by a balance between $I(t)$, the current which leads to the build-up of charge in the cell, and linear pumping, which removes charge from the cell. 
Again, the precise form of $I(t)$ is not important, only two key properties\footnote{The assumption (\ref{epsilon}) was motivated by some numerical experiments with the LRd model \cite{Luo-R}.  Specifically, we found that continued rapid stimulation, which leads to shorter APD's, also increases the concentration of sodium ions in the cell.  If the model is modified by artificially fixing the Na$^+$ concentration at its resting value, the APD's are shortened to a lesser degree \cite{oliver-WKK}. Condition (\ref{epsilon}) is designed to mimic the fast sodium current, which flows at the start of the action potential and only as long as needed to depolarize the cell.  Although such considerations motivated our assumptions, we regard the good fit of the model with experimental data as stronger supporting evidence.}:
\begin{itemize}
\item	$I(t)$ is nonzero only during the upstroke of an action potential (i.e., the initial rapid increase of $v$), and 
\item \e{epsilon}  	\int_{\mathcal{J}} I(t) dt = \epsilon ,		\ee
\end{itemize}
where $\mathcal{J}$ denotes the interval between two consecutive stimuli and  $\epsilon$ is the same-named parameter in (\ref{c}).
In words, (\ref{epsilon}) asserts that a fixed charge $\epsilon$ enters the cell during each action potential.

Under the assumption that
\e{asym-ineq2}
	\tau_{\rsub{in}} \quad \ll \quad \tau_{\rsub{out}}  \quad  \ll \quad \tau_{\rsub{fclose}}, 
	\tau_{\rsub{sclose}}, \tau_{\rsub{open}}  \qquad  \ll \qquad  \tau_{\rsub{pump}} ,
\ee
the mapping (\ref{A}, \ref{c}) may be extracted as an asymptotic limit of (\ref{dvdt-new}, \ref{dhdt-new}, \ref{dcdt}).

The significance of the parameters 
$\tau_{\rsub{pump}}, \tau_{\rsub{fclose}}, \tau_{\rsub{sclose}}$, and $\tau_{\rsub{open}}$ 
in (\ref{A}--\ref{Phi}) is clear from the discussion of the ionic model.
$A_{\rsub{max}}$ is the longest possible action potential since in (\ref{G},\ref{Phi}), the terms
$$
	\tau_{\rsub{fclose}} \ln \left\{ 1- \alpha e^{-D/\tau_{\rsub{open}}} \right\}, \qquad
	\tau_{\rsub{sclose}} \ln \left\{ \frac{e^{-c} + \beta}{1+\beta} \right\}
$$
are nonpositive and tend to zero under repeated slow pacing.
As noted above, $\epsilon$ specifies the amount of charge that enters the cell during each action potential.
The parameter $\alpha$ is a dimensionless combination of several parameters in the ionic model.
Similarly, $\beta$ is the ratio of the strengths of the concentration-independent and concentration-dependent parts of $J_{\rsub{in}}(v,h,c)$.

\section{Fitting the model to experimental data}

The model (\ref{A}--\ref{Phi}) contains eight parameters.
One might attempt to determine all these parameters simultaneously with a massive least-squares fit to the entire restitution portrait. 
However, there probably are multiple local minima of the residual in such a high dimensional parameter space.
Even if there were a unique minimum, there may be a large set in parameter space that gives a fit of essentially the same quality, and in high dimensions it is difficult to assess this prospect quantitatively.
Moreover, such a massive fit would not yield any insight about the significance of the individual parameters.

The restitution portrait provides an attractive alternative---we determine parameters sequentially, in small groups, from four different types of data, as follows:
\begin{itemize}
\item [(1)] Slope of the S1-S2 restitution curve: $ \tau_{\rsub{fclose}}, \alpha$
\item [(2)] Slope of the dynamic restitution curve: 
$\tau_{\rsub{sclose}}, \epsilon \tau_{\rsub{pump}}, \beta$

\item [(3)] Overall height of the dynamic restitution curve: $A_{\rsub{max}}$.
\item [(4)] Transient time constant: $\tau_{\rsub{pump}}$
\end{itemize}	
Note that  $\tau_{\rsub{open}}$ is not on the above list. 
As we shall see below, it turns out that even this sequential approach leaves $\tau_{\rsub{open}}$ quite uncertain; one obtains only a modest lower bound for this parameter.

\subsection{S1-S2 restitution data}	

To determine the slope $S_{12}$  experimentally, many stimuli (say $N$) are applied with a fixed basic cycle length (say  $B$) until the system achieves a steady state, after which the $(N+1)^{st}$ stimulus is applied following a perturbed cycle length $B+\Delta$: then $S_{12}$ is defined by
\e{S12-expt}
	S_{12} = \frac{A_{N+1} - A_N}{D_N - D_{N-1}}.
\ee
Note that
$$
	A_n = A_{ss} \qquad \mbox{for $n$ large and } \, n \le N,
$$
$$
	D_n = D_{ss} \qquad \mbox{for $n$ large and } \, n < N,
$$
while
$$
	D_N = D_{ss} + \Delta , \qquad \mbox{so that}  \qquad  D_N - D_{N-1} = \Delta.
$$

If, under the above protocol, a sequence of action potentials is determined by the mapping (\ref{A}, \ref{c}), then the associated concentrations also satisfy					$$
	c_n = c_{ss} \qquad \mbox{for $n$ large and } \, n \le N ;
$$					
moreover, since by (\ref{asym-ineq2}), $\; \tau_{\rsub{pump}}$ is so large,
$$
	c_{N+1} \approx c_N.
$$
Thus, the model (\ref{A}, \ref{c}) predicts that
\e{S12-th}
	S_{12} \approx G^{\prime}(D_{ss}).
\ee
Differentiating (\ref{G}), we see that the RHS of (\ref{S12-th}) is given by
\e{Gprime}
	G^{\prime}(D_{ss}) = \frac{\tau_{\rsub{fclose}}}{\tau_{\rsub{open}}}
	\frac{1}{e^{D_{ss} / \tau_{\rsub{open}}}  - \alpha }.
\ee

The squares in Figure 1 show $S_{12}$ as obtained in one of the experiments of \cite{kalb-DTIKG}.
It is natural to try to choose the three parameters in (\ref{Gprime}), 
$\tau_{\rsub{fclose}}, \tau_{\rsub{open}}$, and $\alpha$, by finding the best fit to this data.
However, the quality of the fit is not at all sensitive to $\tau_{\rsub{open}}$, provided this parameter is sufficiently large.
This is illustrated in Figure 2, which shows the residual as a function of $\tau_{\rsub{open}}$ when the corresponding optimal values of $\tau_{\rsub{fclose}}$ and $\alpha$ are chosen.
Somwhat  arbitrarily, in the remainder of the fitting process we have chosen a minimal value for this parameter,
\e{tau}
	\tau_{\rsub{open}} = 300 \; \mbox{msec}.
\ee
The corresponding optimal values for $\tau_{\rsub{fclose}}$ and $\alpha$ are given in Table 1, and the fit of the S1-S2 data from the experiment for this value of $\tau_{\rsub{open}}$ is given in Figure~1.
Of course, because the parameters are coupled in various ways, the ambiguity in $\tau_{\rsub{open}}$ means that most of the parameters cannot be regarded as uniquely determined by the experimental data.

\begin{figure}
  \begin{center}
    \epsfig{file = 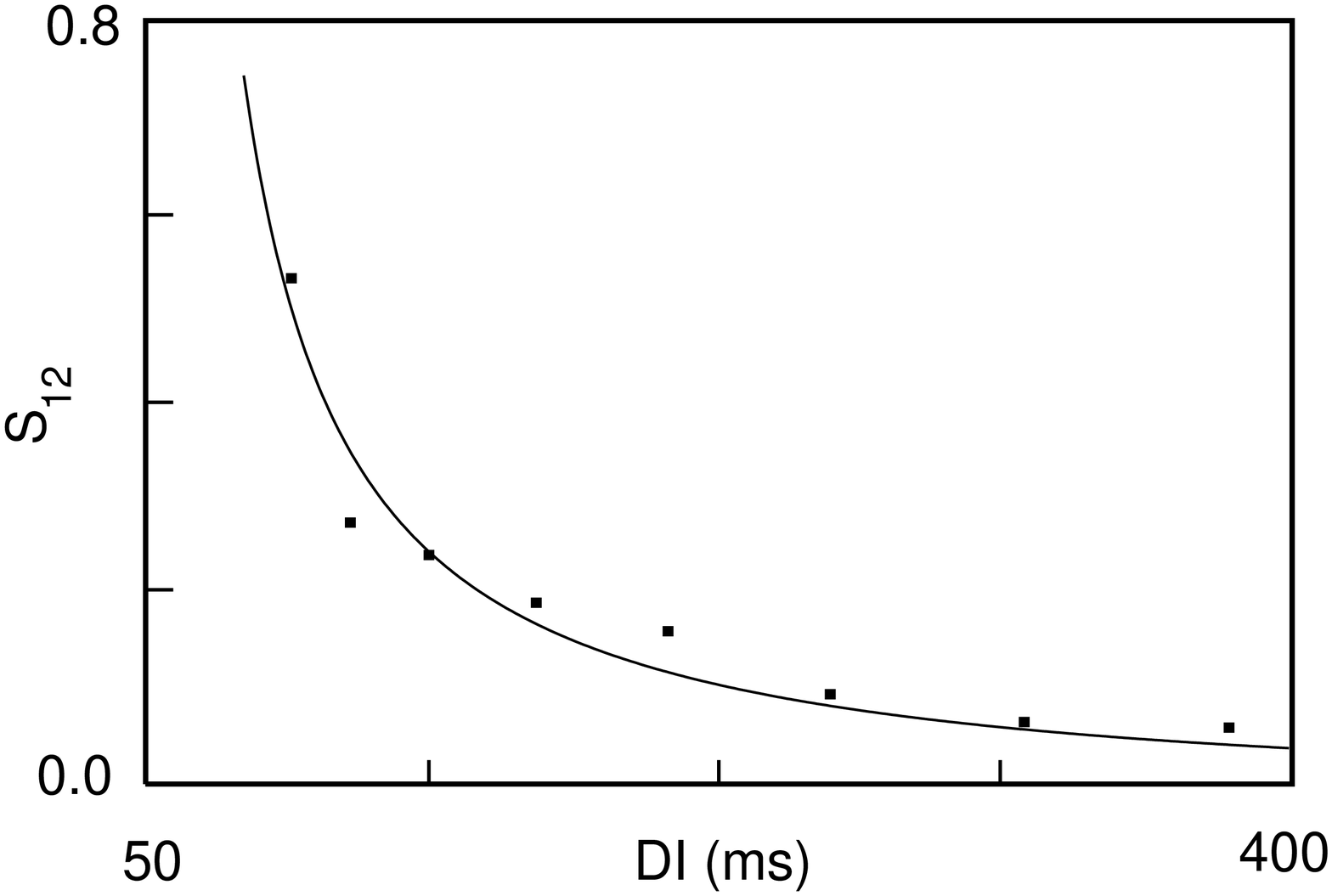, height = 2.0in, width = 3.0in}
   \end{center}
   \caption{A fit of (\ref{Gprime}) to $S_{12}$ data, with $\tau_{\rsub{open}} =$ 300 msec}
\end{figure}

\begin{figure}
  \begin{center}
    \epsfig{file = 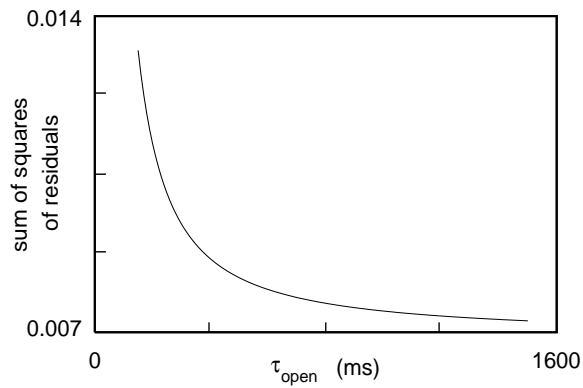, height = 2.0in, width = 3.0in}
   \end{center}
   \caption{The residual in fitting (\ref{Gprime}) to $S_{12}$ data, as a function of $\tau_{\rsub{open}}$}
\end{figure}

\subsection{Slope of the dynamic restitution curve}

According to (\ref{A}), the steady-state action potential duration is given by
\e{Ass1}
	A_{ss} = G(D_{ss}) + \Phi(c_{ss}).
\ee
In the physiological range, $B$ is larger than, but on the order of, a typical action potential. 
The order of magnitude of the action potential duration in the model is determined by the gate constants $\tau_{\rsub{fclose}}, \tau_{\rsub{sclose}}, \tau_{\rsub{open}}$. 
But by (\ref{asym-ineq2}), the gate constants are much smaller than $\tau_{\rsub{pump}}$.
Thus, we conclude that, in the physiological range, 
\e{asym-ineq3}
	B \ll \tau_{\rsub{pump}}.
\ee
Given (\ref{asym-ineq3}), it follows from (\ref{c}) that
\e{css}
	c_{ss} \approx \epsilon \tau_{\rsub{pump}} / B .
\ee
Combining (\ref{Ass1}, \ref{css}), we find that
\e{Ass2}
	A_{ss} = G(D_{ss}) + \Phi(\epsilon \tau_{\rsub{pump}} / B).
\ee
Since $B = A_{ss} + D_{ss}, \; $ (\ref{Ass2}) defines $A_{ss}$ implicitly as a function of $D_{ss}$, and (\ref{Ass2}) may be differentiated to yield the slope of the dynamic restitution curve
\e{Sdyn}
	S_{dyn} = \frac{dA_{ss}}{dD_{ss}} = G^{\prime}(D_{ss}) 
	- \frac{d\Phi}{dc} \; \frac{\epsilon \tau_{\rsub{pump}}}{B^2} \; (S_{dyn} + 1) ,
\ee
where we have used the fact that $dB/D_{ss} =  S_{dyn} + 1$.
Recalling (\ref{S12-th}), computing $d\Phi/dc$ from (\ref{Phi}),  and rearranging, we obtain
\e{Sdyn-fit}
	\frac{S_{dyn} - S_{12}}{S_{dyn} + 1} = \frac{\tau_{\rsub{sclose}} \epsilon \tau_{\rsub{pump}}}{B^2} \;
	\frac{1}{1 + \beta e^{\epsilon \tau_{\rsub{pump}} / B}} .
\ee

The squares in Figure 3 show data for the combination of the slopes $S_{dyn}$ and $S_{12}$ on the LHS of (\ref{Sdyn-fit}) as obtained in one of the experiments\footnote{More accurately, data for $S_{dyn}$ were taken from the experiment, but values for $S_{12}$ from our fit of (\ref{Gprime}) were used.}  
of \cite{kalb-DTIKG}.
The curve in the figure shows the best least-squares fit\footnote{Of the various parameter fits in the paper, Figure 3 is perhaps the least satisfactory.  Note that this weakness is far less clear in the fit of Figure 4, an integrated version of (\ref{Sdyn-fit}).  This difference illustrates the point that data from the restitution portrait provide a demanding test of any mathematical model.}, and the corresponding parameter values for $\tau_{\rsub{sclose}}, \epsilon \tau_{\rsub{pump}}, \beta$ are listed in Table 1.

\begin{table}
\begin{center}
\begin{tabular}{lrc}  \hline
Parameter &				Value &			Units \\  \hline
$ \tau_{\rsub{fclose}}$ &		29.4 & msec \\		
$\alpha$ &				1.174  \\		\hline
$ \tau_{\rsub{sclose}}$ &		570 & msec \\
$\epsilon \tau_{\rsub{pump}}$ &		821 & msec \\
$\beta$ &				0.386 \\		\hline
$\tau_{\rsub{pump}}$ &			30000 & msec \\
$A_{\rsub{max}}$ &			889 & msec \\ 		\hline
\end{tabular}
\end{center}
\caption{Optimal parameter values, assuming $\tau_{\rsub{open}} = 300 \,$ msec}
\end{table}

\begin{figure}
  \begin{center}
    \epsfig{file = 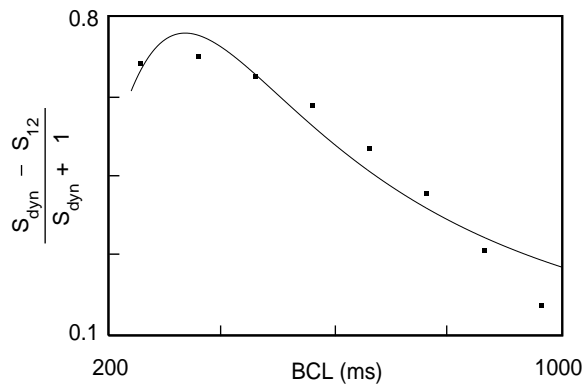, height = 2.0in, width = 3.0in}
   \end{center}
   \caption{A fit of (\ref{Sdyn-fit}) to dynamic restitution slopes}
\end{figure}

\subsection{The two remaining parameters}

We choose $A_{max}$, which sets the overall height of the restitution curve, to fit the data as shown in Figure 4.

Steady state data determines the product $\epsilon \tau_{\rsub{pump}}$ but neither factor individually; for the individual factors, we turn to the transient data.
In the downsweep of the restitution-portrait protocol \cite{kalb-DTIKG}, each time the basic cycle length is decreased, a time constant is extracted from the transient to the next steady state.
These time constants are typically on the order of 20--30 sec.
On the other hand, in the mapping model (\ref{A}, \ref{c}), the time constant of such transients is very nearly equal to $\tau_{\rsub{pump}}$ for all $B$. 
Although the time constants in the experiments vary somewhat with $B$, we merely take $\tau_{\rsub{pump}} = 30,000 \,$ msec,  as indicated in Table 1, which yields $\epsilon = 0.0274$.
(For a possible improvement of this part of the model, see the ``Nonlinear pumping'' discussion in Section~4.3.)

\begin{figure}
  \begin{center}
    \epsfig{file = 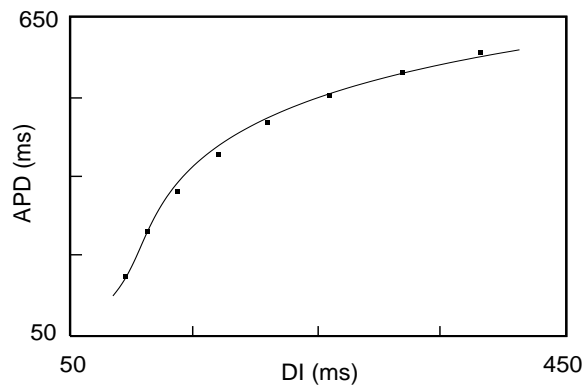, height = 2.0in, width = 3.0in}
   \end{center}
   \caption{A fit of (\ref{A}, \ref{c}) to the dynamic restitution curve}
\end{figure}

\section{Discussion}

\subsection{Conclusions}

In this paper we have introduced an ionically based 2D mapping model for cardiac restitution.
The underlying ionic model suggests some subtle changes from previous models \cite{chialvo-MJ, fox-BG}, and having made these changes, we are able to choose parameters in the mapping to fit S1-S2 restitution data, dynamic restitution data, and most transient data, as obtained in \cite{kalb-DTIKG}.
The simplicity of the procedure means that mathematical descriptions of various experiments could be obtained with these techniques.

We obtained parameters in (\ref{A}, \ref{c}) by fitting different types of data derived from the restitution portrait.
The complete restitution portrait from the experiment \cite{kalb-DTIKG} is shown in Figure~5(a).
As explained in \cite{kalb-DTIKG} (also see the Appendix), such a restitution portrait presents in compact form the dynamic restitution data, the S1-S2 restitution data, and the transient data obtained in the perturbed downsweep protocol.
In Figure~5(b) we show the restitution portrait from the map.
The agreement is remarkably good, with one caveat:
In the experiment, the basic cycle length was decremented in steps of approximately 100 msec.
In computations with the mapping, at the two shortest BCL's, steps this large led to initial diastolic intervals less than the value  $\tau_{\rsub{open}} \ln \alpha$ where (\ref{G}) becomes undefined.
To circumvent this difficulty, it was necessary to decrease the BCL in smaller steps.
Because of this issue, at the two shortest BCL's, the transient shown for the mapping is shorter than for the experiment.
(Remark: In the experiment, data from late in the transients is not recorded and hence is not shown in Figure~5(a).)

In a nonlinear least-squares fitting of data, one wonders whether other, radically different, parameter values could fit the data equally well.
For example, as we saw, the quality of the fit was spectacularly insensitive to $\tau_{\rsub{open}}$.
Accepting the ambiguity with respect to $\tau_{\rsub{open}}$, we further addressed this issue as follows:
In our fitting of (\ref{Gprime}) to $S_{12}$, we evaluated the $2 \times 2$ Hessian matrix of the residual with respect to $\tau_{\rsub{fclose}}$ and $\alpha$;
all diagonal entries of this matrix were large; one of its eigenvalues was also quite large, but the other was rather small.
The situation for fitting (\ref{Sdyn-fit}) was similar, with one even smaller eigenvalue.
From this we conclude that, with $\tau_{\rsub{open}}$ given, changing any one other parameter would affect the quality of the fit adversely, but \emph{groups} of parameters might be changed with little effect.
The issue is still under investigation, in particular how it might affect the transients at shorter basic cycle length.

\begin{figure}
  \begin{center}
    \epsfig{file = 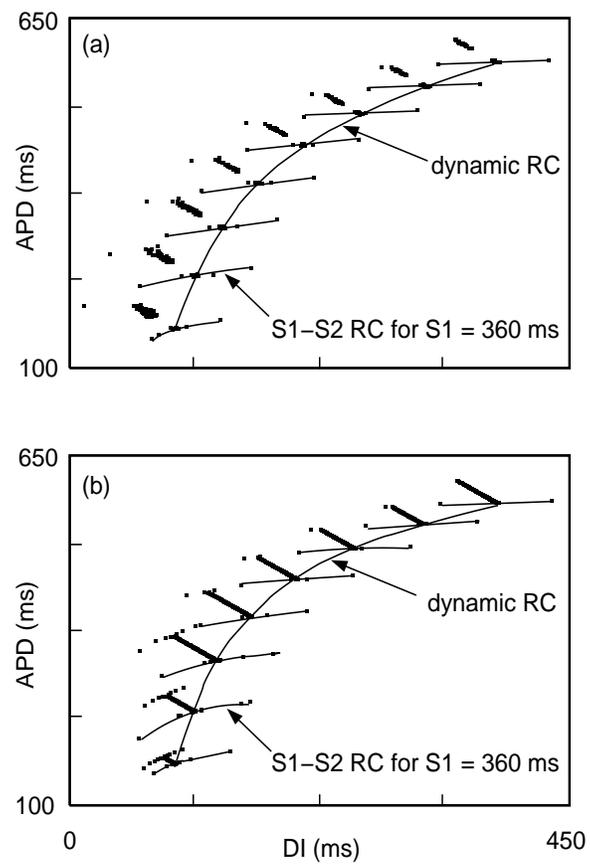, height = 4.5in, width = 3.0in}
   \end{center}
   \caption{The full restitution portrait. (a) Experiment (b) The mapping (\ref{A}, \ref{c})}
\end{figure}

\subsection{Comparison with previous models}

The memory variable $M_n$ in \cite{chialvo-MJ, fox-BG} evolves according to 
\e{Jalife}
	M_{n+1} = \left\{ 1 - (1-M_n) e^{-A_n / \tau} \right\} e^{-D_n / \tau} .
\ee
This equation differs only slightly from (\ref{c}), but this small change was crucial in obtaining a model that exhibited both Behavior~1 and Behavior~2, which were seen in experiments.
Equation (\ref{c}) results from our assumption (\ref{epsilon}) that a fixed charge $\epsilon$ enters the cell during each action potential.
The evolution (\ref{Jalife}) can also be obtained from an ionic model: specifically, a model in which charge accumulates during the action potential and is pumped out during the diastolic interval.

In \cite{nolasco-D, fox-BG}, the analogue of $G(D)$ in (\ref{G}) is given by 
\e{ND}
	K_1 - K_2 e^{-D/ \tau}
\ee
where $K_1, K_2$, and $\tau$ are constants.
The shape of (\ref{ND}) is qualitatively similar to that of (\ref{G}), with 
$K_1, K_2$, and $\tau$ in (\ref{ND}) playing approximately the same role as
$A_{\rsub{max}}, \tau_{\rsub{fclose}}$, and $\tau_{\rsub{open}}$, respectively.
However, (\ref{G}) contains the fourth parameter $\alpha$, which is crucial in fitting the S1-S2 restitution data of \cite{kalb-DTIKG}. 
Incidentally, the logarithmic dependence in (\ref{G}) is what leads to the fact that the memory effect in (\ref{A}) is \emph{additive}, not multiplicative as in \cite{chialvo-MJ, fox-BG}.
Since (\ref{G}) is based on an ionic model, we regard the additive form as more natural, but either form could be fitted to the data.

In \cite{elharrar-S}, Elharrar and Surawicz parametrize the dependence of the steady-state action potential $A_{ss}$ on the period $B$ with a function that has the asymptotic form as $B \rightarrow \infty$
\e{B-infty}
	A_{ss} \sim A_{max} - \mbox{Const} / B .
\ee
This algebraic decay is important since it matches the experimental observation that $A_{ss}$ continues to increase significantly as $B$ is raised far beyond physiological values; by contrast, the exponential decay in (\ref{ND}) would not lead to this behavior.
Let us show that (\ref{A}, \ref{c}) exhibits such algebraic decay for large $B$.
Since $G$ decays exponentially, it will contribute a constant term to (\ref{B-infty}) but no term at the level $\mathcal{O}(1/B)$.  
Recalling that (\ref{asym-ineq3}) holds in the physiological range and referring to (\ref{G}, \ref{Phi}), we obtain (\ref{B-infty}) on substitution into (\ref{Ass1}), where the coefficient of $1/B$ equals 
$$
	\frac{\tau_{\rsub{sclose}} \epsilon \tau_{\rsub{pump}}}{ 1+\beta}.
$$

\subsection{Future work}

\noindent \emph{(a) Alternans}

In this paper we have considered only 1:1 responses of the heart.
Because of the model's transparancy, it is not difficult to adjust parameters in (\ref{A}, \ref{c}) to produce a 2:2 response, or alternans.
However, this does not seem to lead to a good model for the experiments \cite{kalb-DTIKG}, for the following reason:

A mapping leads to alternans when an eigenvalue, say $\lambda$, of its Jacobian at the fixed point passes through -1.  
Because $\tau_{\rsub{pump}}$ is so large, the concentration $c$ has little influence on the relevant eigenvalue; specifically
\e{alt}
\lambda \approx - S_{12}.
\ee
However, it is common in the experiments to find alternans even when $S_{12}$ is on the order of 0.5.
We hope to derive a better model for alternans by including calcium fluxes, as in \cite{shiferaw-WGWK}.

\bigskip

\noindent \emph{(b) Nonlinear pumping}

In our ionic model, (\ref{dcdt}) describes a linear pump that pumps at a very slow rate.
For this model, the time constant for the transient leading to steady state is very nearly equal to $\tau_{\rsub{pump}}$, no matter what the basic cycle length $B$ is. 
However, experiments (e.g., Figure 4 of \cite{elharrar-S}) show that this time constant decreases if $B$ is decreased.
Our model could be modified in several ways to better reproduce this behavior.
Perhaps the simplest is to allow nonlinear pumping: i.e., to change (\ref{dcdt}) to
$$
	\frac{dc}{dt} = I(t) - k c^p 
$$
for some power $p$.
This will be developed in a future publication.

\bigskip

\noindent \emph{(c) Global S1-S2 restitution}

In S1-S2 experiments, after steady state has been attained with stimuli spaced at an interval $S1$, the APD following a perturbed stimulus interval $S2$ is measured.
By a \emph{global} S1-S2 restitution curve, we mean the graph of action potential duration as a function of $S2$, when $S2$ varies over its entire range (with $S1$ fixed). 
By contrast, the restitution portrait 
\cite{kalb-DTIKG} studies only what might be called \emph{local} S1-S2 restitution: i.e., only perturbed intervals $S2$ nearly equal to $S1$ are explored.
Of course global restitution provides more data to test a model, and experiments are planned to test the predictions of the model (\ref{A}, \ref{c}) in this regard.

One positive conclusion may already be reported.
Elharrar and Surawicz \cite{elharrar-S} found that each global S1-S2 curve crosses the dynamic restitution curve twice, both for $S2=S1$ and again for small DI. 
Observe from (\ref{G}) that $G(D)$ tends to negative infinity as $D$ tends to 
$\tau_{\rsub{open}} \ln \alpha > 0$ from above.
Thus, the S1-S2 and dynamic restitution curves produced by (\ref{A}, \ref{c}) also intersect twice in this way.

\bigskip

Support of the National Institutes of Health under grant 1R01-HL-72831 and
the National Science Foundation under grants PHY-0243584 and DMS-9983320 
is gratefully acknowledged.


 \newpage

\section*{Appendix A}

In this Appendix, we briefly summarize the notation used throughout the paper
and the protocols used to generate the various restitution curves.

\subsection*{Tissue response}

When a small piece of cardiac muscle is subjected to a sequence of brief
electrical stimuli who strength exceeds a critical threshold, the myocytes
respond by producing action potentials. \ Figure 6 shows a schematic of
the transmembrane
voltage measured from a single myocyte. \ 
Following a stimulus, first the voltage rises rapidly (indicating cell 
depolarization), then it has a plateau region during which
the cell cannot be reactivated (the refractory period), and finally the voltage returns to its resting value (the cell repolarizes).
This time course is known as an action potential, and the duration of the $n^{th}$ action potential is denoted by $A_{n}$.
The interval between the time when the cell repolarizes (following the $n^{th}$ action potential) and the time when it
depolarizes again (due to the $(n+1)^{st}$ stimulus) is known as the diastolic
interval and is denoted by $D_{n}$.

For a periodic train of stimuli delivered at a slow rate, the myocytes
display a phase-locked 1:1 response, where each stimulus produces an
identical action potential. \ At faster pacing rates, the 1:1 response
pattern is sometimes replaced by a 2:2 phase-locked period-2 response
pattern, known as alternans. \ Under other conditions, either the 1:1 or 2:2
response becomes unstable and is replaced by a 2:1 response pattern, where
only every other stimulus elicits an action potential.

\begin{figure}
  \begin{center}
    \epsfig{file = 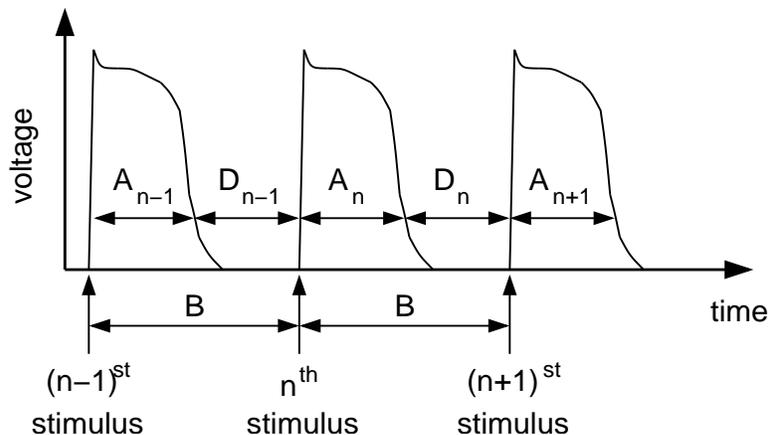, height = 2.25in, width = 4.0in}
   \end{center}
   \caption{Schematic action potentials}
\end{figure}

\subsection*{Electrical restitution}

A fundamental characteristic of cardiac cells is the shortening of $A_{n}$
as the pacing rate increases, known as electrical restitution. \ A
restitution curve quantifies the functional relation between $D_{n}$
and $A_{n}$. \ As pointed out by Elharrar and Surawicz \cite{elharrar-S}, the restitution curve depends
on the pacing protocol used in its measurement. \ Several protocols are
currently in use today, as described in detail in \cite{kalb-DTIKG} and briefly below.

\medskip
{\bf Dynamic restitution curve protocol}

\noindent Pace the tissue periodically with period $B$ until the tissue reaches
steady state. \ Record $A_{n+1}\equiv A^{\ast }$ and $D_{n}\equiv D^{\ast }$%
. \ Change $B$ and repeat. \ The pairs of points $(D^{\ast },A^{\ast })$ form the dynamic restitution curve.

\medskip

{\bf S1-S2 restitution curve protocol}

\noindent Pace the tissue periodically with period $B$ (the ``S1'' stimuli in ``S1-S2'') until the
tissue reaches steady state. \ Make a sudden change in the pacing period
(the ``S2'' stimulus), record the diastolic
interval $D_{n}$ preceding the S2 stimulus and the resulting action
potential duration $A_{n+1}$. \ Repeat for various S2 intervals. \ The pairs
of points $(D_{n},A_{n+1})$ form one of the S1-S2 restitution curves. 
There is a different S1-S2 restitution curve for each value of S1, a characteristic of rate-dependent restitution.

\medskip

{\bf Restitution portrait}

\noindent The restitution portrait is a combination of both protocols\footnote{In this paper we do not consider the constant-B protocol of \cite{kalb-DTIKG}.} described above,
where, for a given $B$, only segments of the S1-S2 restitution curves in
a neighborhood of the corresponding steady state $(D^{\ast },A^{\ast })$ are measured.
The restitution portrait also shows the pairs $(D_{n},A_{n+1})$ in the transient leading to each steady state.

\end{document}